\documentclass[final]{elsarticle}

\usepackage{lineno,hyperref}
\usepackage{amsmath}
\usepackage{amssymb}
\usepackage{subcaption}
\usepackage{color}

\begin{document}

\begin{frontmatter}

\title{Analysis of air pollution time series using complexity-invariant distance and information measures}

\author[unil]{Federico Amato\corref{mycorrespondingauthor}}
\cortext[mycorrespondingauthor]{Corresponding author}
\ead{federico.amato@unil.ch}

\author[unil]{Mohamed Laib}

\author[unil]{Fabian Guignard}

\author[unil]{Mikhail Kanevski}

\address[unil]{IDYST, Faculty of Geosciences and Environment, University of Lausanne, Switzerland}

\begin{abstract}
Air pollution is known to be a major threat for human and ecosystem health. A proper understanding of the factors generating pollution and of the behavior of air pollution in time is crucial to support the development of effective policies aiming at the reduction of pollutant concentration. 
This paper considers the hourly time series of three pollutants, namely NO$_2$, O$_3$ and PM$_{2.5}$, collected on sixteen measurement stations in Switzerland. The  air pollution  patterns due to the location of measurement stations and their relationship with anthropogenic activities, and specifically land use, are studied using two approaches: Fisher-Shannon information plane and complexity-invariant distance between time series. A clustering analysis is used to recognize within the measurements of a same pollutant groups of stations behaving in a similar way. The results  clearly demonstrate the relationship between the air pollution probability densities and land use activities.
\end{abstract}


\end{frontmatter}

\section{Introduction}

Air pollution is one of the major global threats causing deep impacts on human health and ecosystems \cite{franck2011effect, Peled2011}. It is considered to be the major cause of premature death and disease, and the International Agency for Research on Cancer has classified it as carcinogenic \cite{IARC2013, pope2006health}. Particulate Matter (PM) is known to have huge impacts on human health, being related to respiratory problems, bronchitis, reduced lung functionality \cite{who2003health, martuzevicius2004spatial}.

Air pollution has strong impact on flora and fauna, along with soil and water quality. Most of the ecosystem services are nowadays threatened by the exposure to air pollutant \cite{he2001characteristics}. Nitrogen dioxide (NO$_2$) interferes with both terrestrial and marine ecosystems, as it increases the quantity of nutrient nitrogen in the natural systems causing eutrophication and potentially leading to changes in species diversity or to invasion of new species \cite{driscoll2003nitrogen}. Moreover, NO$_2$ is partly responsible for the acidification of soils and waters. Ground level ozone (O$_3$) emissions can impact agricultural crops as well as forests as they reduce the growth rate of plants, hence having negative impacts on biodiversity and ecosystem services \cite{wang2007ground}.

Climate change is strongly related to pollutant concentration too. Both O$_3$ and some of the constituent of PM such as black carbon are short-term forcers directly contributing to global warming. Moreover, the already mentioned impact of O$_3$ on vegetation growth may also cause a reduction of its capacity of uptake of carbon dioxide. On the other hand, climate change may enlarge the impacts on the environment of several air pollutant \cite{hansen2011climate}.

Many other spheres of human life are impacted by air pollution. Indeed, it can cause damage to the built environment and to cultural heritage through the biodegradation and soiling, caused by the particulate matter, or the fading of colours inducted by O$_3$. Finally, air pollution also have negative economic impacts, mainly due to the reduced labour productivity, additional health expenditure and crops or yield losses \cite{oced, TRIPPETTA2013105}. 

The recognition of the effects of air pollution has led the international community to the development of an increasing amount of policies to reduce the emissions of pollutants. Air pollution is nowadays perceived as one of the biggest concerns by citizens \cite{lorenzoni2006public}, and major efforts have been done by policy makers to meet the World Health Organization air quality guidelines \cite{krzyzanowski2008update}, as well as the United Nations Sustainable Development Goals. The latter policy framework targets to substantially reduce the number of deaths and illness caused by air pollution by 2030, and improving air quality plays a key role in reaching this target\cite{assembly2015sustainable, griggs2013policy}.

Modelling and interpreting air pollution data are recognized as key steps to foster the reduction of the negative effects of air pollution \cite{kumar2018five}. Several studies have been paying attention to the characterization of air pollution time series \cite{belis2013critical, YATKIN2007126, salcedo1999time}, their estimations with statistical learning methods \cite{Alimissis2018}, and to the understanding of the pollution sources \cite{lin2018extreme, elangasinghe2014complex, Huang2019}. Moreover, in \cite{TELESCA20084387, telesca2009analysis, telesca2011complexity} authors used complexity measures to characterise the dynamics of several pollutant, including cadmium, iron, lead and particulate matter.

In this study the time series of three pollutants, namely NO$_2$, O$_3$ and PM$_{2.5}$, collected with hourly frequency on sixteen measurement stations in Switzerland are analyzed. Focus of the paper is on the characterization of the  considered time series using complexity measures to provide insights concerning the phenomena and  pollution patterns due to the location of measurement stations and its relationship with anthropogenic activities and land use. Specifically, a  clustering analysis will be performed by applying  a complexity-invariant distance between time series  to recognize within the measurements of a same pollutant groups of stations behaving in a similar way. The discovered clusters will then be represented in the Fisher-Shannon information plane  for a deeper understanding of the  air pollution patterns. The applied  methods will give the results in line with the physical behaviour of the pollution phenomenon, enabling for a better understanding of the patterns of air pollution, their dependencies on land use and the level of predictability. These contributions are expected to be useful to support the development of more efficient policies aiming at the reduction of emission of pollutant in air.

The remainder of the paper is organized as follows. Section \ref{M&M} explores the air pollution dataset investigated in this paper and presents the methodological framework for the application of time series clustering and for the construction of the Fisher-Shannon plane. Section \ref{R&D} presents and discusses the results obtained. Finally, section \ref{C} summarizes the conclusions and discusses potential future works.

\section{Materials and Methods}\label{M&M}

\subsection{Data collection}
This paper explores the relationship between the behaviour of several air pollutant and land use. The latter is here considered to be an indicator of the anthropogenic local activities in the area. Hourly pollution time series have been collected from sixteen measuring station belonging to the Swiss National Air Pollution Monitoring Network (NABEL) \cite{fur2010technischer}. 

\begin{figure}
\includegraphics[width=\linewidth]{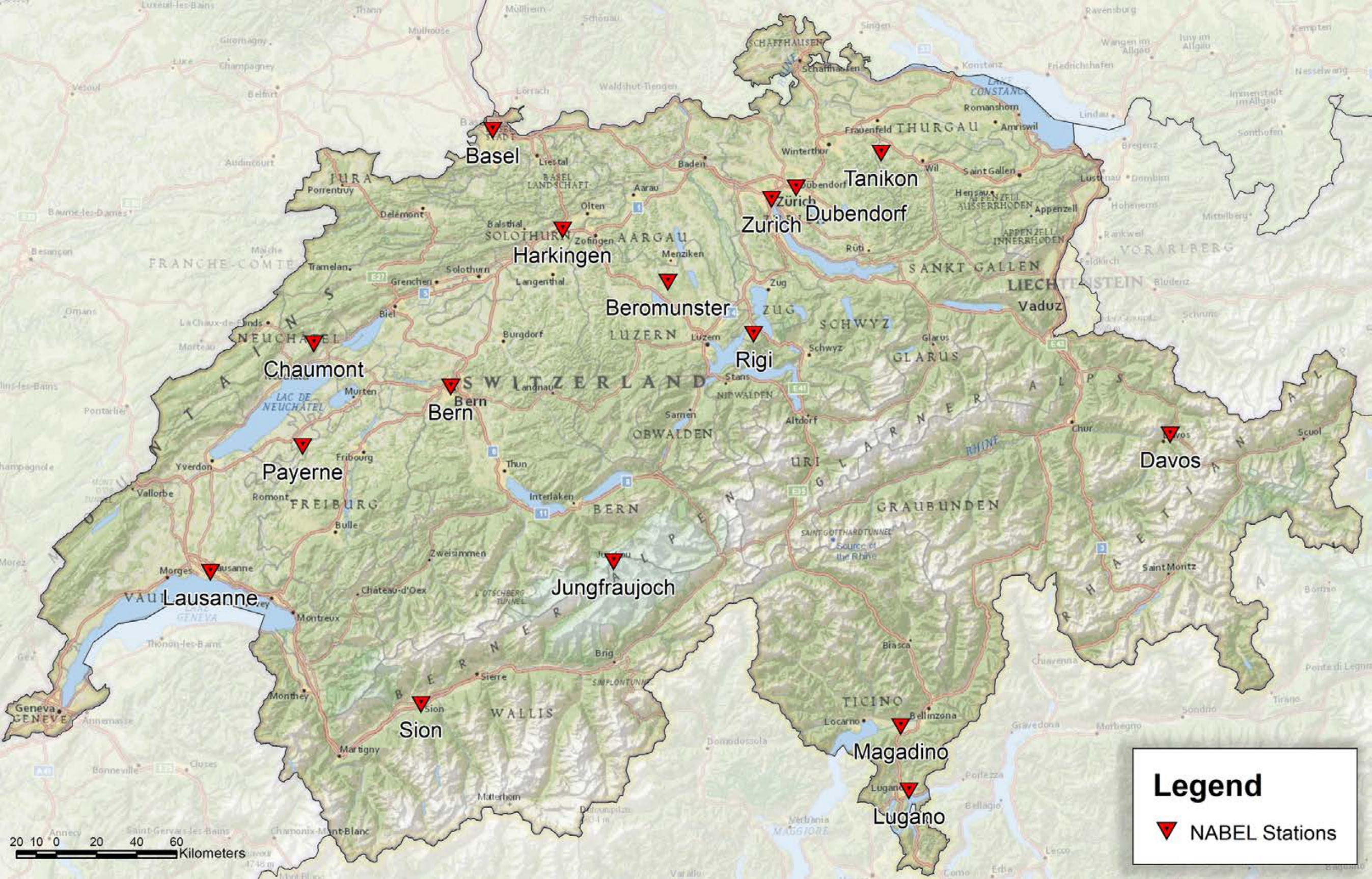}
\caption{The location of the sixteen measuring station belonging to the NABEL network in Switzerland.}
\label{NABEL_stations}  
\end{figure}

Three different pollutants have been studied, namely NO$_2$, O$_3$ and PM$_{2.5}$. The stations are located in different environments including urban, suburban, rural and mountainous sites as well as along highways (Fig. \ref{NABEL_stations}).  The location of the stations is extremely relevant, because we will consider the corresponding land use as an indication of the type of human activities in the neighbouring area of the stations (Table \ref{staloc}).

\begin{table}
\begin{center}
\begin{tabular}{ |c|l|l|c|c|c| } 
\hline
ID & Station &  Location (land use) & PM$2.5$ & $O_3$ & $NO_2$\\ 
\hline
1 & Bern & Urban-traffic & $\surd$ & $\surd$ & $\surd$ \\
2 & Lausanne & Urban-traffic & -  & $\surd$ &  $\surd$\\
3 & Zurich & Urban & $\surd$ & $\surd$ &  $\surd$ \\
4 & Lugano & Urban & $\surd$ & $\surd$ &  $\surd$ \\
5 & Basel & Suburban & $\surd$ & $\surd$ &  $\surd$ \\
6 & Dubendorf & Suburban & $\surd$ & $\surd$ &  $\surd$ \\
7 & Harkingen & Rural-Highway & $\surd$ & $\surd$ &  $\surd$ \\
8 & Sion & Rural-Highway & -  & $\surd$ &   $\surd$\\
9 & Beromunster & Rural $<$ $1000$ m  & $\surd$ & $\surd$ &  $\surd$ \\
10 & Magadino & Rural $<$ $1000$ m  & $\surd$ & $\surd$ &  $\surd$ \\
11 & Payerne & Rural $<$ $1000$ m  & $\surd$ & $\surd$ &  $\surd$ \\
12 & Tanikon & Rural $<$ $1000$ m  & - & $\surd$ &   $\surd$ \\
13 & Chaumont & Rural $>$ $1000$ m  & -  &  $\surd$&   $\surd$\\
14 & Davos & Rural $>$ $1000$ m  & - &  $\surd$&   $\surd$ \\
15 & Rigi & Rural $>$ $1000$ m  & $\surd$ & $\surd$ &  $\surd$ \\\
16 & Jungfraujoch & High-montain &  $\surd$ &  $\surd$ & - \\
\hline
\end{tabular}
\caption{Location of the measuring station. for each station, the land use in the station area and the measured pollutant are indicated.}
\label{staloc}
\end{center}
\end{table}

\begin{figure}
\centering
\includegraphics[width=1\linewidth]{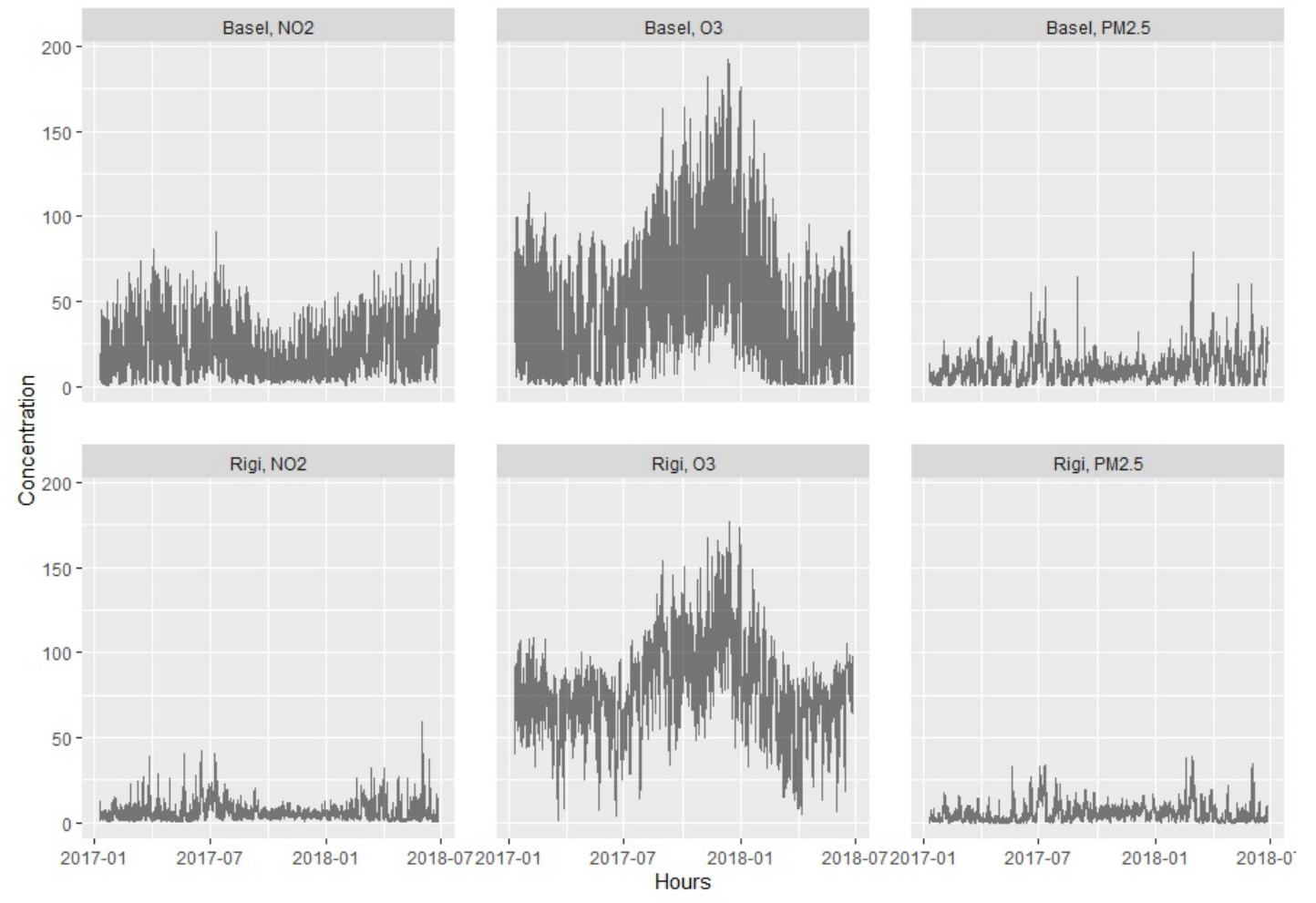}
\caption{Examples of the time series of  NO$_2$, O$_3$ and PM$_{2.5}$ measured at the Basel and Rigi stations.}
\label{TSexample}  
\end{figure}

The data were recorded from 1st of September 2017 to 28th of February 2019. Figure \ref{TSexample} shows examples of the time series plots of the three pollutants used in this research. A complete exploratory data analysis is provided in the Supplementary Materials, highlighting how the distribution of the pollutants concentration are skewed with right tail, showing the presence of extreme values in the time series and temporal (cross)correlations. The time series exhibit different cycles, including a 12 hours one and a weekly one. However, only the daily cycle was detectable for all the time series. Hence, we only removed it from the raw data, and we considered the remaining cyclical effects as a significant part of the physical process of the emission of pollutant to be studied. Daily cycle and trend have been removed by Seasonal Decomposition of Time Series by Loess (STL) \cite{cleveland1990stl}. Hereafter, the research focuses on the remainder of the time series after the STL decomposition in order to extract information locally.

\subsection{Fisher-Shannon plane}

A random variable can be described by two positive real numbers - the Shannon Entropy Power (SEP) and the Fisher Information Measure (FIM)\cite{telesca2013fisher}. Let $X$ be a univariate continuous random variable following a Probability Density Function (PDF), noted $f(x)$. The SEP of $X$, noted $N_X$, is defined in \cite{Shannon1948, Cover2006}  

\begin{equation}\label{N}
N_X = \frac{1}{2\pi e} \exp\{2H_X\},
\end{equation}
where $H_X$ is the differential entropy provided by
$
H_X = \mathbb E \left[\log f(x)\right].
$

The FIM of $X$, noted $I_X$, is also known as the Fisher information of $X$ with respect to a scalar translation parameter \cite{Dembo1991}. It is defined as follows \cite{Cover2006} 
\begin{equation}\label{I}
I_X = \mathbb E\left[ \left(\frac{\partial }{\partial x} \log f(x)\right)^2\right].
\end{equation}

The PDF of $X$ can then be analyzed displaying the SEP and FIM within the so-called Fisher-Shannon (FS) plane \cite{Vignat2003, telesca2011analysis, lovallo2011complexity}. In the FS plane, the only reachable values are in the set $\{(N_X, I_X) \in \mathbb R^2 | N_X >0, \, I>0 \text{ and } N_X \cdot I_X \geq 1 \}$. This is due to the isoperimetric inequality for entropies which states that $N_X \cdot I_X \geq 1$, with equality if and only if X is a Gaussian random variable \cite{Dembo1991, Cover2006}.

Integrals were numerically estimated  to obtain the SEP and the FIM from data. More precisely, $f(x)$ and its derivative $f'(x)$ are replaced by their kernel density estimators (KDE) in the integral forms of \eqref{N} and \eqref{I}, \cite{Bhattacharya1967, Dmitriev1973, PrakasaRao1983, Gyorfi1987, Joe1989}.
The KDE approach is the following \cite{Wand1994}: given $n$ independent realizations $\{x_1, \dots, x_n\}$ of $X$, the PDF $f(x)$ is approximated by 
\begin{equation}\label{KDE}
    \hat f_h(x) = \frac{1}{nh}\sum_{i=1}^n K\left( \frac{x-x_i}{h} \right),
\end{equation}
where $h$ is the bandwidth parameter and $K(u)$ is the kernel, which is supposed to be a unimodal probability density function that is symmetric around zero and has an integral over $\mathbb R$ summing up to 1.

In this paper, the Gaussian kernel with zero mean and unit variance is used, and the estimator \eqref{KDE} becomes
\begin{equation}\label{KDEGauss}
    \hat f_h(x) = \frac{1}{nh\sqrt{2\pi}}\sum_{i=1}^n \exp\left\{\frac{1}{2}\left( \frac{x-x_i}{h} \right)^2\right\}.
\end{equation}
A natural estimates for the PDF derivative, $f'(x)$, is obtained by deriving $\hat f_h(x)$, which yields
\begin{equation}\label{KDEderivGauss}
    \hat f'_h(x) = \frac{1}{nh^3 \sqrt{2\pi}}\sum_{i=1}^n (x-x_i)\exp\left\{\frac{1}{2}\left( \frac{x-x_i}{h} \right)^2\right\}.
\end{equation}
Finally, the estimate of \eqref{N} and \eqref{I} are respectively
\[
\hat N_X = \frac{1}{2\pi e}\exp\left\{2\int \hat f_h(x) \log \hat f_h(x) \, dx \right\},
\]
and
\[
\hat I_X = \int \frac{\left(f'_{h}(x)\right)^2}{f_h(x)} \, dx.
\]
The bandwidth is chosen by the Sheather-Jones direct plug-in method \cite{Sheather1991}, which approximates the optimal bandwidth with respect to the Asymptotic Mean Integrated Squared Error (AMISE).
The quantity $\hat N_X $ measures the amount of disorder  in the data, while $\hat I_X$ is a measure which quantifies its organization.

\subsection{Time series clustering}\label{TSclustering}
Time series clustering refers to the unsupervised problem of assigning labels to unlabeled data objects \cite{montero2014tsclust}. Specifically, two time series are assigned to the same group if they are highly similar according to a predefined criterion. Once a similarity measure is computed between the time series, clustering is performed using conventional algorithms. Among these, one of the most popular is \textit{k-}means \cite{liao2005clustering}.

Crucial point in time series clustering is the choice of a suitable similarity/dissimilarity measure. There are many measures and features proposed in time series clustering, see, for example a review in \cite{ReviewClustering} . In this paper the Complexity-Invariant Distance measure is adopted (CID) \cite{batista2014cid}. CID is based on the use of a measure of complexity difference between two time series as a correction factor for a standard distance measure between two time series. Let $X$ and $Y$ be two time series. 
\\
A simple complexity estimate can be defined as:

\begin{equation}
    CE_X = \sqrt{\sum_{i=1}^{n-1}(x_i-x_{i+1})^2}.
\end{equation}

By using an euclidean distance, the Complexity-Invariant Distance measure is computed as follows:

\begin{equation}\label{CID}
    CID_{(X,Y)}=CF_{(X,Y)} \sqrt{\sum_{i=1}^n(x_i-y_i)^2} 
\end{equation}
where $CF_{(X,Y)}$ is the correction factor defined by
\begin{equation}
    CF_{(X,Y)}=\frac{\max(CE_X, CE_Y)}{\min(CE_X, CE_Y)}.
\end{equation}

Once the distance between time series has been computed, clustering can be performed using \textit{k-}means. Evidently, the problem of the selection of the optimal number of clusters has to be solved \cite{kodinariya2013review}. To this aim, in the application proposed in this paper the silhouette width has been used \cite{rousseeuw1987silhouettes}. Silhouette measures how well an object - in this case a time series - fits to its own cluster. The silhouette width for the $i-th$ object is defined as:
\begin{equation}
    s(i) = \frac{b(i)-a(i)}{\max(a(i), b(i))}
\end{equation}
in which $a(i)$ is the average distances between $i$ and the other objects belonging to the same cluster to which $i$ is attributed, and $b(i)$ is the minimum between the average distances among the object $i$ and all the objects in the other clusters. 

A set of clusters can be characterized by the average silhouette width. A common procedure to select the optimal number of clusters in \textit{k-}means is to test different values of \textit{k} and select as optimal the number of clusters resulting in the higher average silhouette width.

\section{Results and Discussion}\label{R&D}

To analyze the behaviour of the three pollutants studied in this paper, $\hat I_X$ and $\hat N_X $ have been computed. Subsequently, for each pollutant, a clustering analysis has been carried out. The results have then been plotted in the Fisher-Shannon plane to show the relationship of the order/disorder of the pollutant time series with the land use information.

\begin{figure}
\includegraphics[width=.5\linewidth]{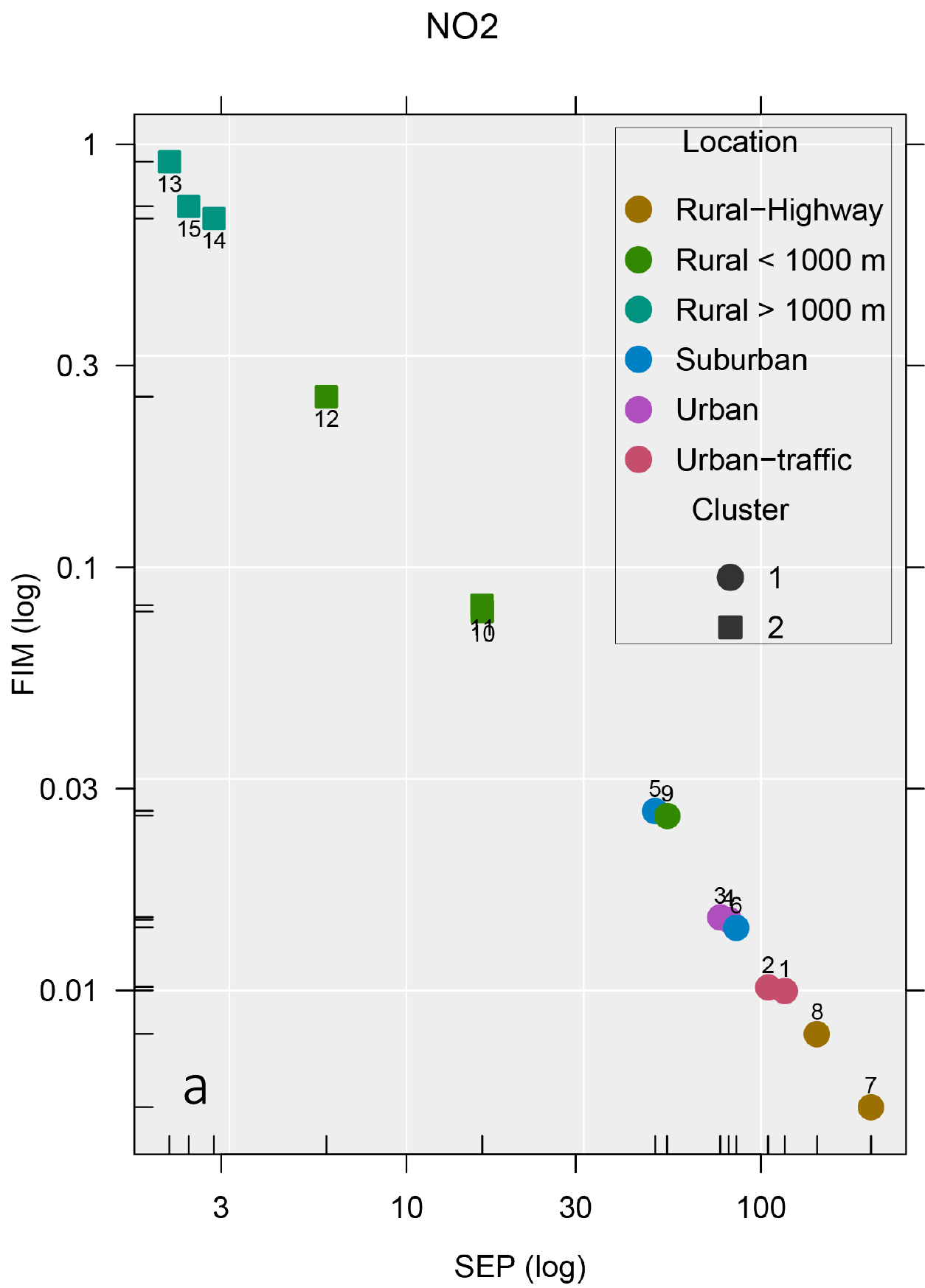}
\includegraphics[width=.5\linewidth]{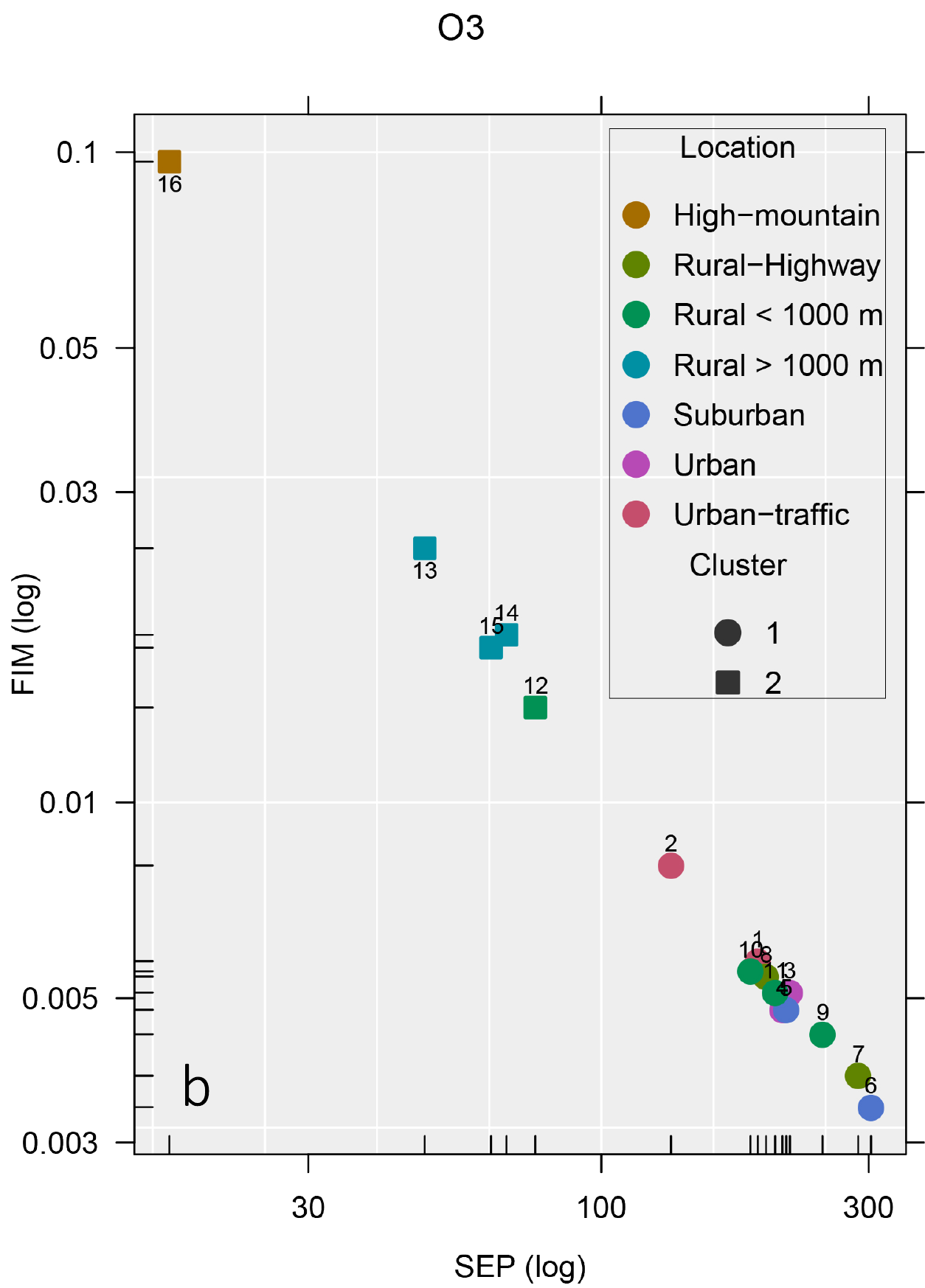}
\includegraphics[width=.5\linewidth]{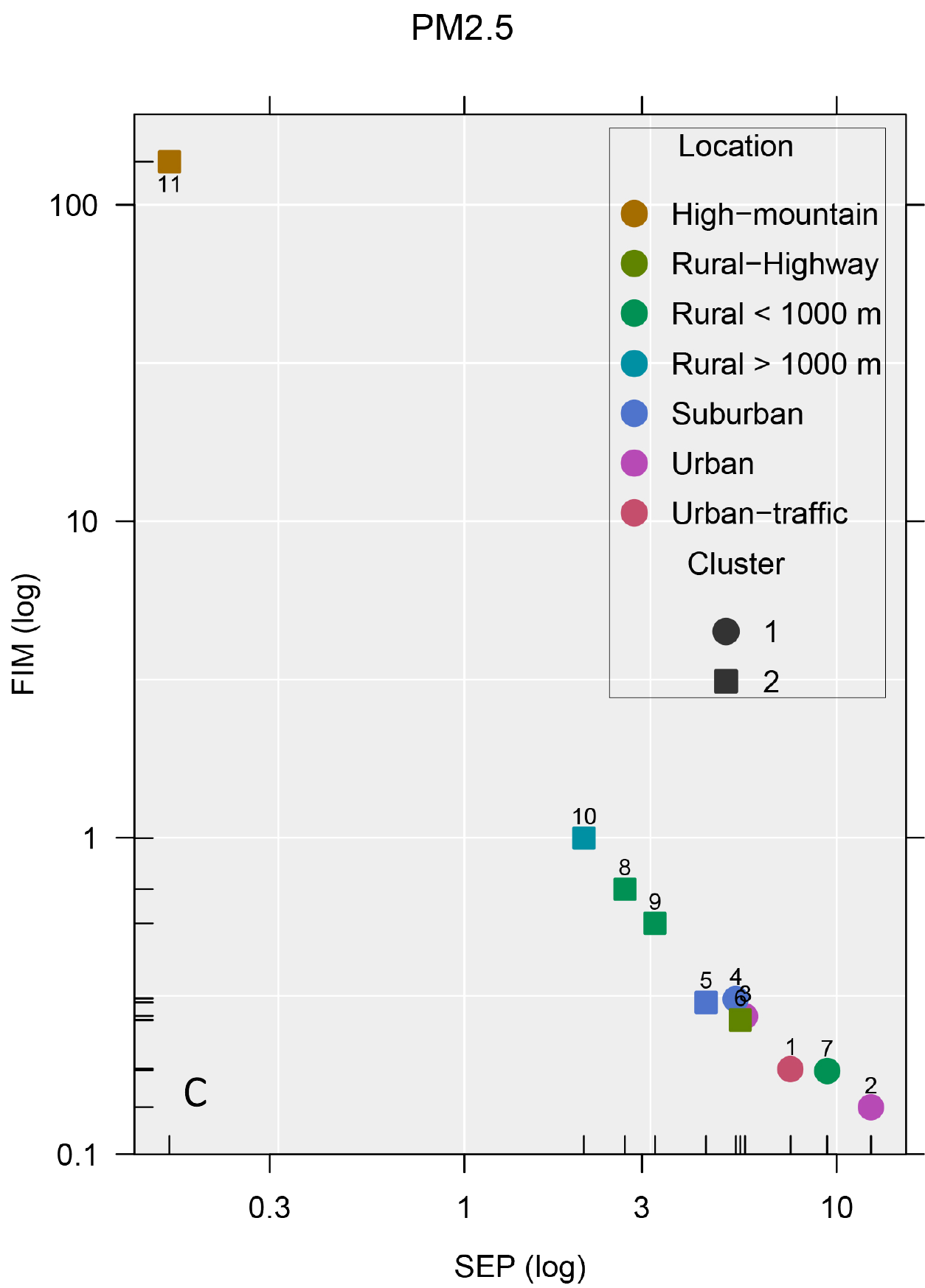}
\includegraphics[width=.5\linewidth]{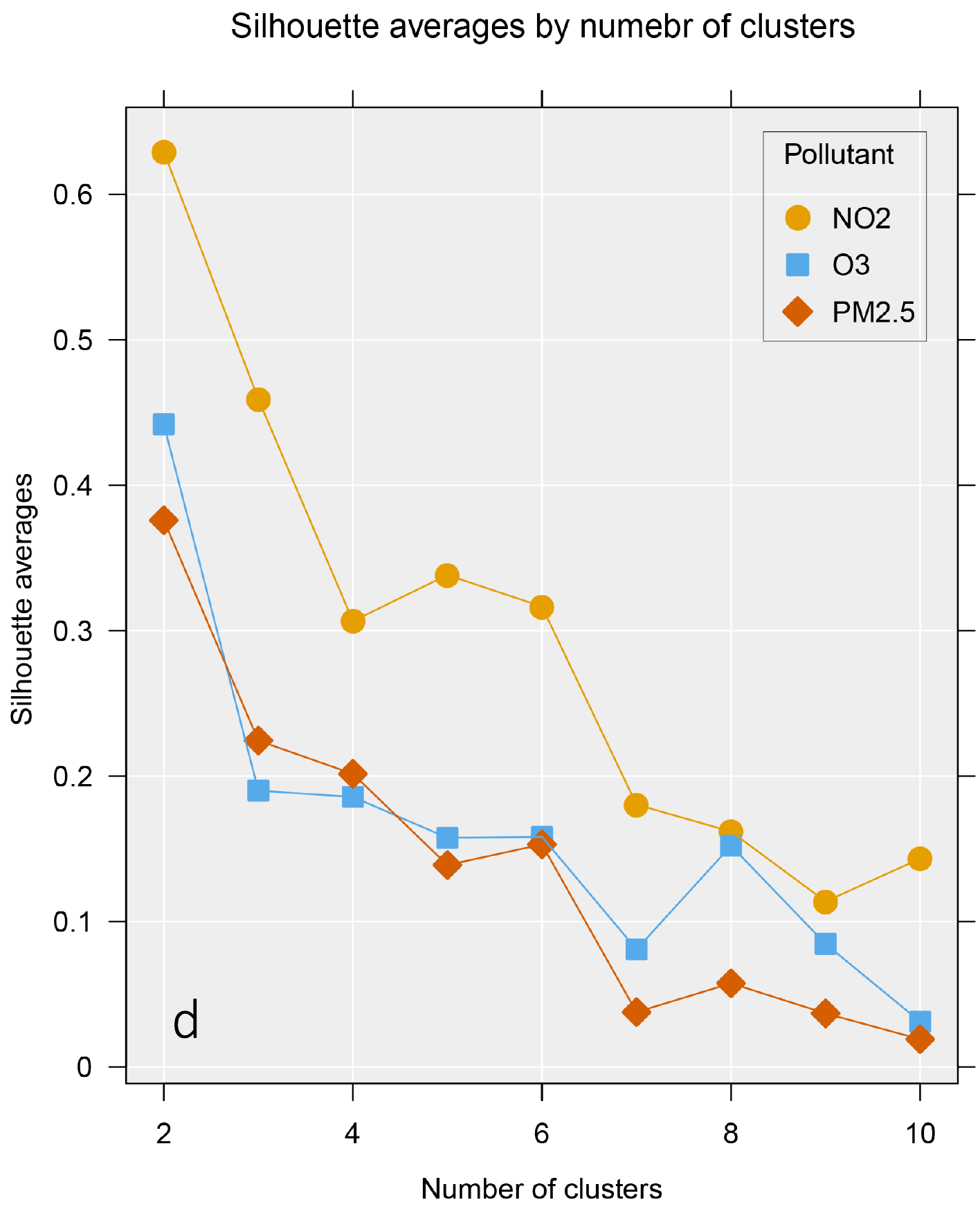}
\caption{Fisher-Shannon plane for NO$_2$ (\textbf{a}), O$_3$ (\textbf{b}) and PM$_{2.5}$ (\textbf{c}) time series. Axis are in log-scale. The numeric labels are the same as in Table \ref{staloc}. Points are colored according to the location of each station. while their shape indicate the cluster to which they belong. 
\\
(\textbf{d}): Average silhouette width for different number of clusters and for each pollutant. In the three cases, the optimal number of clusters is found to be equal to two.
}
\label{FSplane}  
\end{figure}

Both $\hat I_X$ and $\hat N_X $ have been computed for NO$_2$, O$_3$ and PM$_{2.5}$ at all the measurement stations. Figure \ref{FSplane} shows the Fisher-Shannon plane for the three different pollutants, together with the location of each station. The pattern of the points in the planes is extremely interesting. Indeed, it unveils a dependence of the rate of information and disorder on the specific location of each station and, specifically, on the kind of land use in the area. This pattern is extremely clear for NO$_2$ and PM$_{2.5}$. These pollutants are generated as a consequence of combustion of fuels, for power generation, domestic heating or transport reasons. 

The Air Quality in Europe report of 2018 \cite{EuropeanEnvironmentalAgency2018} indicated the road transport as one of the sectors mostly contributing to the total NO$_2$ and PM$_{2.5}$ emission, together with the commercial, institutional and household sectors. Being related to traffic, the emission of the two pollutants can vary significantly during the day, i.e. being higher in the rush hours, or in seasons. Similarly, household emission are basically related to heating, which is mainly needed in winter, or cooling, needed only in the hotter weeks of summer. This irregularity in the emission of the pollutants explains the high degree of disorder measured for the stations located in urban, urban-traffic, suburban and highway sites. For these stations, the highest value of $\hat N_X $ and the lowest  of $\hat I_X$ have been found. Moreover, it is interesting to highlight how in the case of NO$_2$ the stations showing the highest level of disorder are the ones located in the urban-traffic and rural-highway zones. Indeed, \cite{EuropeanEnvironmentalAgency2018} indicates how transports are the major contributors to NO$_2$ pollution, being responsible for the 39\% of its emission. At the same time, the 56\% of PM$_{2.5}$ is generated by commercial, institutional and household activities, represented in our dataset in the classes urban and urban traffic, which have also been found to have the highest  disorder in the corresponding plane. For NO$_{2}$, the Beromunster station, located in a rural area ($<$1000 meters), has a different pattern from the other stations belonging to the same class, showing values of $\hat I_X$ and $\hat N_X $ similar to those of suburban areas. Further analysis should be done on this station in order  to characterize this behaviour properly.

Up to now, O$_3$ has not been mentioned. That is because it is usually not directly emitted in atmosphere, but is generated as a reaction between NO$_x$, hydrocarbons and sunlight. A well known consequence of this chemical rule is the so-called ozone paradox. Ozone levels tend to be as high in rural areas than they are in cities, as in the latter there is a higher availability of NO, generally produced by traffic, that can react with O$_3$ to generate nitrogen dioxide and oxygen. The degree of disorder measured for the different stations, is, therefore, somehow a reflection of the availability of NO at the station area. As a consequence, in the O$_3$ Fisher-Shannon plane the stations of Beromunster, Magadino and Payerne have a behaviour close to those of urban stations.

For each pollutant, time series have been clustered using the approach described in Section \ref{TSclustering}. Figure \ref{FSplane}-d shows the average silhouette values for the three pollutants, highlighting how for the three of them the optimal number of clusters is two. These two clusters are represented in the Fisher-Shannon plane (Figure \ref{FSplane} a-c). Once again, the relationship with the land use and human activities in the location of the stations is recovered, showing how the group including the stations in rural areas have the highest values of Fisher Information and the lowest of Shannon Entropy Power, indicating low degree of disorder. Conversely, the second cluster includes the stations which time series have a higher level of disorder, having an higher level of uncertainty and a lower level of information. Relevant patterns, like those described for O$_3$ rural stations, alre also recovered by the clusters.
Further investigations should analyze whether this pattern has also implications on the level of predictability of the phenomenon. 

\section{Conclusions}\label{C}

It this paper hourly air pollution time series of NO$_2$, O$_3$ and PM$_{2.5}$ from sixteen stations in Switzerland have been analyzed. Fisher Information Measure and Shannon Entropy Power have been computed for each time series, allowing the recognition of a clear relationship between the level of disorder in the concentration of air pollutant and the location of the measurement station, which has been treated as an indicator of the type of land use and of the human activities in the area. This permitted to relate the behaviour of the time series to their emission sources. A clustering analysis further supported the results by the identification of two different groups of time series, the one collected in the traffic-related or urban areas and the one collected in the rural sites.

The data collected at the locations characterized by the presence of stronger anthropogenic sources of pollutant emission have been found to have greater degrees of disorder, with higher values of Shannon Entropy and lower values of Fisher Information. This statistical finding is in line with the physical behaviour of the pollution phenomenon, which is known to be dependent from time-irregular or yearly-seasonal emissions. However, the higher disorder of data collected in urban areas or along trafficked highways may also imply a lower level of predictability of the emissions. This is an extremely relevant finding, because of the emission forecasts are considered to be crucial to support the development of effective policies to reduce the air pollution, especially in urban areas.

\section{Acknowledgements} 

F. Guignard and M. Kanevski acknowledge the support of the National Research Programme 75 “Big Data” (PNR75, project No. 167285 "HyEnergy") of the Swiss National Science Foundation (SNSF).

M. Laib thanks the support of “Soci\'et\'e Acad\'emique Vaudoise” (SAV) and the Swiss Government Excellence Scholarships.

\bibliography{mybibfile}

\begin{thebibliography}{10}
\expandafter\ifx\csname url\endcsname\relax
  \def\url#1{\texttt{#1}}\fi
\expandafter\ifx\csname urlprefix\endcsname\relax\def\urlprefix{URL }\fi
\expandafter\ifx\csname href\endcsname\relax
  \def\href#1#2{#2} \def\path#1{#1}\fi

\bibitem{franck2011effect}
U.~Franck, S.~Odeh, A.~Wiedensohler, B.~Wehner, O.~Herbarth, The effect of
  particle size on cardiovascular disorders—the smaller the worse, Science of
  the Total Environment 409~(20) (2011) 4217--4221.

\bibitem{Peled2011}
R.~Peled,
  \href{https://www.sciencedirect.com/science/article/pii/S1352231011000033}{{Air
  pollution exposure: Who is at high risk?}}, Atmospheric Environment 45~(10)
  (2011) 1781--1785.
\newblock \href {http://dx.doi.org/10.1016/J.ATMOSENV.2011.01.001}
  {\path{doi:10.1016/J.ATMOSENV.2011.01.001}}.
\newline\urlprefix\url{https://www.sciencedirect.com/science/article/pii/S1352231011000033}

\bibitem{IARC2013}
IARC, {Outdoor air pollution a leading environmental cause of cancer deaths},
  Tech. rep., International Agency for Research on Cancer, WHO (2013).

\bibitem{pope2006health}
C.~A. Pope~III, D.~W. Dockery, Health effects of fine particulate air
  pollution: lines that connect, Journal of the air \& waste management
  association 56~(6) (2006) 709--742.

\bibitem{who2003health}
W.~WHO, Health aspects of air pollution with particulate matter, ozone and
  nitrogen dioxide, World Health Organization Working Group Bonn, Germany,
  13-15.

\bibitem{martuzevicius2004spatial}
D.~Martuzevicius, S.~A. Grinshpun, T.~Reponen, R.~L. G{\'o}rny, R.~Shukla,
  J.~Lockey, S.~Hu, R.~McDonald, P.~Biswas, L.~Kliucininkas, et~al., Spatial
  and temporal variations of pm2. 5 concentration and composition throughout an
  urban area with high freeway density—the greater cincinnati study,
  Atmospheric Environment 38~(8) (2004) 1091--1105.

\bibitem{he2001characteristics}
K.~He, F.~Yang, Y.~Ma, Q.~Zhang, X.~Yao, C.~K. Chan, S.~Cadle, T.~Chan,
  P.~Mulawa, The characteristics of pm2. 5 in beijing, china, Atmospheric
  Environment 35~(29) (2001) 4959--4970.

\bibitem{driscoll2003nitrogen}
C.~T. Driscoll, D.~Whitall, J.~Aber, E.~Boyer, M.~Castro, C.~Cronan, C.~L.
  Goodale, P.~Groffman, C.~Hopkinson, K.~Lambert, et~al., Nitrogen pollution in
  the northeastern united states: sources, effects, and management options,
  BioScience 53~(4) (2003) 357--374.

\bibitem{wang2007ground}
X.~Wang, W.~Manning, Z.~Feng, Y.~Zhu, Ground-level ozone in china: distribution
  and effects on crop yields, Environmental pollution 147~(2) (2007) 394--400.

\bibitem{hansen2011climate}
L.~J. Hansen, J.~R. Hoffman, Climate savvy: Adapting conservation and resource
  management to a changing world, Island Press, 2011.

\bibitem{oced}
OECD,
  \href{https://www.oecd-ilibrary.org/content/publication/9789264257474-en}{The
  Economic Consequences of Outdoor Air Pollution} (2016).
\newblock \href
  {http://dx.doi.org/https://doi.org/https://doi.org/10.1787/9789264257474-en}
  {\path{doi:https://doi.org/https://doi.org/10.1787/9789264257474-en}}.
\newline\urlprefix\url{https://www.oecd-ilibrary.org/content/publication/9789264257474-en}

\bibitem{TRIPPETTA2013105}
S.~Trippetta, R.~Caggiano, L.~Telesca,
  \href{http://www.sciencedirect.com/science/article/pii/S1352231013003567}{Analysis
  of particulate matter in anthropized areas characterized by the presence of
  crude oil pre-treatment plants: The case study of the agri valley (southern
  italy)}, Atmospheric Environment 77 (2013) 105 -- 116.
\newblock \href
  {http://dx.doi.org/https://doi.org/10.1016/j.atmosenv.2013.05.013}
  {\path{doi:https://doi.org/10.1016/j.atmosenv.2013.05.013}}.
\newline\urlprefix\url{http://www.sciencedirect.com/science/article/pii/S1352231013003567}

\bibitem{lorenzoni2006public}
I.~Lorenzoni, N.~F. Pidgeon, Public views on climate change: European and usa
  perspectives, Climatic change 77~(1-2) (2006) 73--95.

\bibitem{krzyzanowski2008update}
M.~Krzyzanowski, A.~Cohen, Update of who air quality guidelines, Air Quality,
  Atmosphere \& Health 1~(1) (2008) 7--13.

\bibitem{assembly2015sustainable}
G.~Assembly, Sustainable development goals, SDGs), Transforming our world: the
  2030.

\bibitem{griggs2013policy}
D.~Griggs, M.~Stafford-Smith, O.~Gaffney, J.~Rockstr{\"o}m, M.~C. {\"O}hman,
  P.~Shyamsundar, W.~Steffen, G.~Glaser, N.~Kanie, I.~Noble, Policy:
  Sustainable development goals for people and planet, Nature 495~(7441) (2013)
  305.

\bibitem{kumar2018five}
R.~Kumar, V.-H. Peuch, J.~H. Crawford, G.~Brasseur, Five steps to improve
  air-quality forecasts (2018).

\bibitem{belis2013critical}
C.~Belis, F.~Karagulian, B.~R. Larsen, P.~Hopke, Critical review and
  meta-analysis of ambient particulate matter source apportionment using
  receptor models in europe, Atmospheric Environment 69 (2013) 94--108.

\bibitem{YATKIN2007126}
S.~Yatkin, A.~Bayram,
  \href{http://www.sciencedirect.com/science/article/pii/S0169809506002882}{Elemental
  composition and sources of particulate matter in the ambient air of a
  metropolitan city}, Atmospheric Research 85~(1) (2007) 126 -- 139.
\newblock \href
  {http://dx.doi.org/https://doi.org/10.1016/j.atmosres.2006.12.002}
  {\path{doi:https://doi.org/10.1016/j.atmosres.2006.12.002}}.
\newline\urlprefix\url{http://www.sciencedirect.com/science/article/pii/S0169809506002882}

\bibitem{salcedo1999time}
R.~Salcedo, M.~A. Ferraz, C.~Alves, F.~Martins, Time-series analysis of air
  pollution data, Atmospheric Environment 33~(15) (1999) 2361--2372.

\bibitem{Alimissis2018}
A.~Alimissis, K.~Philippopoulos, C.~Tzanis, D.~Deligiorgi,
  \href{https://www.sciencedirect.com/science/article/pii/S1352231018305119}{{Spatial
  estimation of urban air pollution with the use of artificial neural network
  models}}, Atmospheric Environment 191 (2018) 205--213.
\newblock \href {http://dx.doi.org/10.1016/J.ATMOSENV.2018.07.058}
  {\path{doi:10.1016/J.ATMOSENV.2018.07.058}}.
\newline\urlprefix\url{https://www.sciencedirect.com/science/article/pii/S1352231018305119}

\bibitem{lin2018extreme}
C.~Lin, R.-J. Huang, D.~Ceburnis, P.~Buckley, J.~Preissler, J.~Wenger,
  M.~Rinaldi, M.~C. Facchini, C.~O’Dowd, J.~Ovadnevaite, Extreme air
  pollution from residential solid fuel burning, Nature Sustainability 1~(9)
  (2018) 512.

\bibitem{elangasinghe2014complex}
M.~Elangasinghe, N.~Singhal, K.~Dirks, J.~Salmond, S.~Samarasinghe, Complex
  time series analysis of pm10 and pm2. 5 for a coastal site using artificial
  neural network modelling and k-means clustering, Atmospheric Environment 94
  (2014) 106--116.

\bibitem{Huang2019}
Z.~Huang, Q.~Yu, W.~Ma, L.~Chen,
  \href{https://www.sciencedirect.com/science/article/pii/S1352231019305047}{{Surveillance
  efficiency evaluation of air quality monitoring networks for air pollution
  episodes in industrial parks: Pollution detection and source
  identification}}, Atmospheric Environment (2019) 116874\href
  {http://dx.doi.org/10.1016/J.ATMOSENV.2019.116874}
  {\path{doi:10.1016/J.ATMOSENV.2019.116874}}.
\newline\urlprefix\url{https://www.sciencedirect.com/science/article/pii/S1352231019305047}

\bibitem{TELESCA20084387}
L.~Telesca, R.~Caggiano, V.~Lapenna, M.~Lovallo, S.~Trippetta, M.~Macchiato,
  \href{http://www.sciencedirect.com/science/article/pii/S0378437108002483}{The
  fisher information measure and shannon entropy for particulate matter
  measurements}, Physica A: Statistical Mechanics and its Applications 387~(16)
  (2008) 4387 -- 4392.
\newblock \href {http://dx.doi.org/https://doi.org/10.1016/j.physa.2008.02.064}
  {\path{doi:https://doi.org/10.1016/j.physa.2008.02.064}}.
\newline\urlprefix\url{http://www.sciencedirect.com/science/article/pii/S0378437108002483}

\bibitem{telesca2009analysis}
L.~Telesca, R.~Caggiano, V.~Lapenna, M.~Lovallo, S.~Trippetta, M.~Macchiato,
  Analysis of dynamics in cd, fe, and pb in particulate matter by using the
  fisher--shannon method, Water, air, and soil pollution 201~(1-4) (2009)
  33--41.

\bibitem{telesca2011complexity}
L.~Telesca, M.~Lovallo, Complexity analysis in particulate matter measurements,
  Computational Ecology and Software 1~(3) (2011) 146.

\bibitem{fur2010technischer}
S.~B. f{\"u}r Umwelt, E.~K{\"o}rperschaft, Technischer bericht zum nationalen
  beobachtungsnetz f{\"u}r luftfremdstoffe (nabel) 2010, Tech. rep., ETH Zurich
  (2010).

\bibitem{cleveland1990stl}
R.~B. Cleveland, W.~S. Cleveland, J.~E. McRae, I.~Terpenning, Stl: a
  seasonal-trend decomposition, Journal of official statistics 6~(1) (1990)
  3--73.

\bibitem{telesca2013fisher}
L.~Telesca, M.~Lovallo, Fisher-shannon analysis of wind records, International
  Journal of Energy and Statistics 1~(04) (2013) 281--290.

\bibitem{Shannon1948}
C.~E. Shannon, A mathematical theory of communication, Bell System Technical
  Journal 27~(3) (1948) 379--423.
\newblock \href {http://dx.doi.org/10.1002/j.1538-7305.1948.tb01338.x}
  {\path{doi:10.1002/j.1538-7305.1948.tb01338.x}}.

\bibitem{Cover2006}
T.~M. Cover, J.~A. Thomas, Elements of Information Theory (Wiley Series in
  Telecommunications and Signal Processing), Wiley-Interscience, New York, NY,
  USA, 2006.

\bibitem{Dembo1991}
A.~Dembo, T.~M. Cover, J.~A. Thomas, Information theoretic inequalities, IEEE
  Transactions on Information Theory 37~(6) (1991) 1501--1518.
\newblock \href {http://dx.doi.org/10.1109/18.104312}
  {\path{doi:10.1109/18.104312}}.

\bibitem{Vignat2003}
C.~Vignat, J.-F. Bercher, Analysis of signals in the fisher–shannon
  information plane, Physics Letters A 312~(1) (2003) 27 -- 33.
\newblock \href
  {http://dx.doi.org/https://doi.org/10.1016/S0375-9601(03)00570-X}
  {\path{doi:https://doi.org/10.1016/S0375-9601(03)00570-X}}.

\bibitem{telesca2011analysis}
L.~Telesca, M.~Lovallo, Analysis of the time dynamics in wind records by means
  of multifractal detrended fluctuation analysis and the fisher--shannon
  information plane, Journal of Statistical Mechanics: Theory and Experiment
  2011~(07) (2011) P07001.

\bibitem{lovallo2011complexity}
M.~Lovallo, L.~Telesca, Complexity measures and information planes of x-ray
  astrophysical sources, Journal of Statistical Mechanics: Theory and
  Experiment 2011~(03) (2011) P03029.

\bibitem{Bhattacharya1967}
P.~K. Bhattacharya, Estimation of a probability density function and its
  derivatives, Sankhyā: The Indian Journal of Statistics, Series A (1961-2002)
  29~(4) (1967) 373--382.

\bibitem{Dmitriev1973}
Y.~Dmitriev, F.~Tarasenko, On the estimation of functionals of the probability
  density and its derivatives, Theory of Probability \& Its Applications 18~(3)
  (1973) 628--633.
\newblock \href {http://dx.doi.org/10.1137/1118083}
  {\path{doi:10.1137/1118083}}.

\bibitem{PrakasaRao1983}
B.~Prakasa~Rao, Nonparametric Functional Estimation, Probability and
  Mathematical Statistics: A Series of Monographs and Textbooks, Academic
  Press, 1983.

\bibitem{Gyorfi1987}
L.~Györfi, E.~C. van~der Meulen, Density-free convergence properties of
  various estimators of entropy, Computational Statistics and Data Analysis
  5~(4) (1987) 425 -- 436.
\newblock \href
  {http://dx.doi.org/https://doi.org/10.1016/0167-9473(87)90065-X}
  {\path{doi:https://doi.org/10.1016/0167-9473(87)90065-X}}.

\bibitem{Joe1989}
H.~Joe, Estimation of entropy and other functionals of a multivariate density,
  Annals of the Institute of Statistical Mathematics 41~(4) (1989) 683--697.
\newblock \href {http://dx.doi.org/10.1007/BF00057735}
  {\path{doi:10.1007/BF00057735}}.

\bibitem{Wand1994}
M.~Wand, M.~Jones, Kernel Smoothing, Chapman \& Hall/CRC Monographs on
  Statistics \& Applied Probability, Taylor \& Francis, 1994.

\bibitem{Sheather1991}
S.~J. Sheather, M.~C. Jones, A reliable data-based bandwidth selection method
  for kernel density estimation, Journal of the Royal Statistical Society.
  Series B (Methodological) 53~(3) (1991) 683--690.

\bibitem{montero2014tsclust}
P.~Montero, J.~A. Vilar, et~al., Tsclust: An r package for time series
  clustering, Journal of Statistical Software 62~(1) (2014) 1--43.

\bibitem{liao2005clustering}
T.~W. Liao, Clustering of time series data —- a survey, Pattern recognition
  38~(11) (2005) 1857--1874.

\bibitem{ReviewClustering}
S.~Aghabozorgi, A.~S. Shirkhorshidi, T.~Y. Wah, Time-series clustering – a
  decade review, Information Systems 53 (2015) 16 -- 38.

\bibitem{batista2014cid}
G.~E. Batista, E.~J. Keogh, O.~M. Tataw, V.~M. De~Souza, Cid: an efficient
  complexity-invariant distance for time series, Data Mining and Knowledge
  Discovery 28~(3) (2014) 634--669.

\bibitem{kodinariya2013review}
T.~M. Kodinariya, P.~R. Makwana, Review on determining number of cluster in
  k-means clustering, International Journal 1~(6) (2013) 90--95.

\bibitem{rousseeuw1987silhouettes}
P.~J. Rousseeuw, Silhouettes: a graphical aid to the interpretation and
  validation of cluster analysis, Journal of computational and applied
  mathematics 20 (1987) 53--65.

\bibitem{EuropeanEnvironmentalAgency2018}
{European Environmental Agency}, {Air quality in Europe - 2018 report},
  Publication Office of the European union, Luxembourg, 2018.
\newblock \href {http://dx.doi.org/10.2800/777411} {\path{doi:10.2800/777411}}.

\end{thebibliography}

\end{document}